\documentclass[
reprint,
showpacs,
nofootinbib,
 amsmath,amssymb,
 aps,
pra,
floatfix,
]{revtex4-1}

\usepackage{graphicx}
\usepackage{dcolumn}
\usepackage{bm}
\usepackage[colorlinks=true, citecolor=blue, urlcolor=blue ]{hyperref}
\usepackage{amsmath, amssymb}
\usepackage{pxfonts,txfonts}

\newcommand{\bq}{\begin{equation}}
\newcommand{\eq}{\end{equation}}
\newcommand{\ba}{\begin{array}}
\newcommand{\ea}{\end{array}}
\newcommand{\bqa}{\begin{eqnarray}}
\newcommand{\eqa}{\end{eqnarray}}

\newcommand{\ran}{\rangle}
\newcommand{\lan}{\langle}

\begin{document}


\title{Tight lower bound on geometric discord of bipartite states}
\author{Swapan Rana}
 \email{swapanqic@gmail.com}
\author{Preeti Parashar}
 \email{parashar@isical.ac.in}
\affiliation{Physics and Applied Mathematics Unit, Indian Statistical Institute, 203 B T Road, Kolkata, India}
\date{\today}

\begin{abstract} We use singular value decomposition to derive a tight lower bound for geometric discord of arbitrary bipartite states. In a single shot this also leads to an upper bound of measurement induced non locality which in turn yields that for Werner and isotropic states the two measures coincide. We also emphasize that our lower bound is saturated for all $2\otimes n$ states. Using this we show that both the generalized $GHZ$ and $W$ states of $N$ qubits satisfy monogamy of geometric discord. Indeed, the same holds for all $N$-qubit pure states which are equivalent to $W$ states under stochastic local operations and classical communication. We show by giving an example that not all pure states of four or higher qubits satisfy monogamy.
\end{abstract}


\maketitle



Recent years have witnessed the emergence of some non classical correlations other than entanglement. Of them, the quantum discord is the most well studied and it indicates that separable states may possess \emph{quantumness} which can be exploited in various tasks e.g., state merging.  There are different versions of quantum discord and their measures. However, almost all measures are very difficult to calculate analytically, except the geometric discord (GD) introduced by Daki\'{c} \emph{et al.} \cite{dakicprl10}. GD is defined as \bq\label{defgd} D(\rho)=\min_{\chi\in\Omega_0}\|\rho-\chi\|^2\eq where $\Omega_0$ is the set of zero-discord states (i.e., \emph{classical-quantum states}, given by $\sum p_k|\psi_k\ran\lan \psi_k|\otimes\rho_k$) and $\|A\|^2=$ Tr$(A^\dagger A)$ is the Frobenius or Hilbert-Schmidt norm. The authors in \cite{dakicprl10} have also calculated $D$ for arbitrary 2-qubit states, using the explicit Bloch representation.  This however poses a problem in generalizing the formula since the explicit Bloch representation is not known beyond 2-qubits (particularly, conditions for a vector $v\in \mathbb{R}^{d^2-1}$ to represent the Bloch vector of a qu-$d$it is not known for $d\ge3$). So, this problem can not be solved analytically, in general. Fortunately, Luo and Fu have given an alternative description of GD in \cite{luofupra10}, via a minimization over all possible von Neumann measurements on $\rho^a$ \bq\label{drhopi}D(\rho)=\min_{\Pi^a}\|\rho-\Pi^a(\rho)\|^2\eq and cast GD as the following
optimization problem:
\bq\label{luogd1} D(\rho)=\text{Tr}(CC^t)-\max_A\text{Tr}(ACC^tA^t)\eq
where $C=(C_{ij})$ is an  $m^2\times n^2$ matrix, given by the expansion \bq\label{exprho}\rho=\sum c_{ij}X_i\otimes Y_j\eq in terms of orthonormal operators $X_i\in L(H^a), Y_j\in L(H^b)$ and $A=(a_{ki})$ is an $m\times m^2$ matrix given by \bq\label{resta}a_{ki}=\text{Tr}|k\ran\lan k|X_i=\lan k|X_i|k\ran\eq for any orthonormal basis $\{|k\ran\}$ of $H^a$. Thus, the problem of determination of $D$ reduces to finding the maximum of $f(A):=$Tr$\left(ACC^tA^t\right)$ subject to the restriction in (\ref{resta}).  Some effort has been directed towards this last part \cite{joag10}. In this Brief Report, we derive a lower bound of GD for
arbitrary states which will be shown to be saturated by all $2\otimes n$ states.

Another \emph{post-entanglement} measure of quantum correlations is the measurement induced nonlocality (MIN), introduced by Luo and Fu \cite{luofuprl11}. The MIN is defined as somewhat dual to the GD, by \bq\label{dmin}N(\rho)=\max_{\Pi^a}\|\rho-\Pi^a(\rho)\|^2\eq
where the maximum is taken over the von Neumann measurements $\Pi^a=\{\Pi_k^a\}$ which do not disturb $\rho^a$ locally, that is \bq\label{dmincond}\sum_k\Pi_k^a\rho^a\Pi_k^a=\rho^a\eq Thus, MIN is an indicator of the \textit{global effect} on the whole system $\rho^{ab}$ caused by \emph{locally invariant measurement} applied to one part, $\rho^a$. In \cite{luofuprl11}, the authors have calculated MIN for arbitrary pure states and $2\otimes n$ mixed states. We will show that our lower bound of GD automatically reduces to the upper bound of MIN derived therein. Using this bound, we show that the Werner and isotropic states have same amount of GD and MIN. These states are good candidates for \textit{maximally entangled} states and have been studied frequently in literature.

\emph{A tight lower bound on geometric Discord for arbitrary states:}
To solve the optimization problem in (\ref{luogd1}), it is helpful to fix the orthonormal bases $\{X_i\},\{Y_j\}$ and usually the following Bloch representation is considered: \bq\label{blochrho}\rho=\frac{1}{mn}\left[I_m\otimes I_n+\mathbf{x}^t\mathbf{\lambda}\otimes I_n+I_m\otimes\mathbf{y}^t\mathbf{\lambda}+\sum T_{ij}\lambda_i\otimes \lambda_j\right]\eq where $\mathbf{\lambda}=(\lambda_1,\lambda_2,\ldots,\lambda_{d^2-1})^t$ with $\lambda_i$ being the generators of $SU(d)$ for appropriate dimension $d=m,n$ \cite{blochvectorsud}. Comparing the two forms of $\rho$ given by (\ref{exprho}) and (\ref{blochrho}), we identify $X_1=\frac{1}{\sqrt{m}}I_m, Y_1=\frac{1}{\sqrt{n}}I_n, X_{i\ne1}=\frac{1}{\sqrt{2}}\lambda_{i-1},Y_{j\ne1}=\frac{1}{\sqrt{2}}\lambda_{j-1}$ and    \bq\label{C}C=\frac{1}{\sqrt{mn}}\left(                                                                                                                                                                \begin{array}{cc}
1 &\sqrt{\frac{2}{n}}\mathbf{y}^t\\
\sqrt{\frac{2}{m}}\mathbf{x} & \frac{2}{\sqrt{mn}}T \\
\end{array}
\right)\eq
Next we observe that the restriction (\ref{resta}) basically gives the following three restrictions on  $A$:
\begin{subequations}
\label{mresta1}
\bqa \mathbf{e}^t:=(a_{k1})_{k=1}^m=(\lan k|X_1|k\ran)_{k=1}^m=\frac{1}{\sqrt{m}}(1,1,\ldots,1)&~~~&\label{conde}\\
\sum_{k=1}^m \mathbf{a_k}:=\sum_{k=1}^m\left(a_{ki}\right)_{i=2}^{m^2}=\left(\sum_{k=1}^ma_{ki}\right)_{i=2}^{m^2}=\left(\text{Tr} X_i\right)_{i=2}^{m^2}=\mathbf{0}&~~~&\label{condcolumn}\\
\text{  the \emph{isometry condition} }\quad AA^t=I_m&~~~&\label{conda}\\
\text{  and  }|k\ran\lan k|\text{ should be a legitimate pure state. }\label{coherentv}\eqa\end{subequations}
Before proceeding further, we note that the condition (\ref{coherentv}) means $(a_{ki})_{i=1}^{m^2}$ should be a \emph{coherence vector} for all $k$ and as mentioned before, there is no known sufficient condition for it beyond $\mathbb{R}^3$. Thus, this constraint generically can not be implemented into the optimization problem for $m\ge3$. So, for the the time being, let us ignore this constraint and optimize (maximize) $f(A)$ with respect to the other constraints. Clearly, that would give us a lower bound of $D(\rho)$.

To incorporate  (\ref{conde}) into $A$, we write $A=(\mathbf{e}~B)$ where $B$ is any $m\times m^2-1$ matrix subject to the restrictions (\ref{condcolumn}) and (\ref{conda}). With these forms of $A$ and $C$, we have
\bqa\label{tracca}
f(A)&=&\frac{1}{mn}
\left[\left(1+\frac{2}{n}\right)\|\mathbf{y}\|^2+2\text{Tr}\left\{B\left(\sqrt{\frac{2}{m}}\mathbf{x}+\frac{2\sqrt{2}}{n\sqrt{m}}T\mathbf{y}
\right)\mathbf{e}^t\right\}\right.\nonumber\\&~&\left.+\text{Tr}\left\{B\left(\frac{2}{m}\mathbf{xx}^t+\frac{4}{mn}TT^t\right)B^t\right\}\right]
        \eqa

Noting that $\mathbf{xe}^t=\frac{1}{\sqrt{m}}(\mathbf{x},\mathbf{x},\ldots,\mathbf{x})$, we have Tr$(B\mathbf{xe}^t)=\sum_{k=1}^m\mathbf{a_k}.\mathbf{x}=0$, by (\ref{condcolumn}). Similarly, noting that $T\mathbf{y}$ is a column vector, we have Tr$(BT\mathbf{ye}^t)=0$ and hence the first trace term in (\ref{tracca}) vanishes. So, we are left with only the second trace term.

Writing $A=(\mathbf{e}~B)$, we have from (\ref{conda}), $(\mathbf{e}~B)(\mathbf{e}^t~B^t)^t=I_m$, or $\mathbf{ee}^t+BB^t=I_m$. Thus $B$ must satisfy \bq\label{restB} BB^t=I_m-\mathbf{ee}^t\eq This shows that the eigenvalues of $BB^t$ are 1 (with multiplicity $m-1$) and $0$ (with $\mathbf{e}$ being an eigenvector). Let us choose an $m\times m^2-1$ orthogonal matrix $U$ having $\mathbf{e}$ as its last column. Then, every $B$ satisfying (\ref{restB}) can be written as $B=U\Sigma V^t$, where $V$ is an $m^2-1\times m^2-1$ orthogonal matrix and $\Sigma$ is an $m\times m^2-1$ diagonal matrix with diagonal $(1,1,\ldots,1,0)$. Then defining $G:=\left(\frac{2}{m}\mathbf{xx}^t+\frac{4}{mn}TT^t\right)$, for brevity, the last term in (\ref{tracca}) becomes
\bqa\label{gB} g(B)&=&\text{Tr}\left[BGB^t\right]=\text{Tr}\left[U\Sigma V^tGV\Sigma^tU^t\right]\nonumber\\
&=&\text{Tr}\left[\Sigma^tU^tU\Sigma V^tGV\right]=\text{Tr}\left[\Delta V^tGV\right]\eqa

where $\Delta:=\Sigma^t\Sigma=\text{diag}(I_{m-1},0_{m^2-m})$. This shows that maximum of $g(B)$ occurs when $V^tGV$ is a diagonal matrix whose diagonal entries are in non-increasing order. Since $G$ is real symmetric, there always exists such an orthogonal $V$. Hence we have\bq\label{finalg}\max g(B)=\sum_{k=1}^{m-1}\lambda_{k}^{\downarrow}\eq where $\lambda_{k}^{\downarrow}$ are the eigenvalues of $G$ sorted in non-increasing order. Substituting this value of g(B) in (\ref{tracca}), we get $\max f(A)$ which in turn gives the desired lower bound for GD from (\ref{luogd1}) as
\bq\label{finaldiscord} D(\rho)\ge\frac{1}{mn}\left[\frac{2}{m}\mathbf\|{x}\|^2+\frac{4}{mn}\|T\|^2-\sum_{k=1}^{m-1}\lambda_k^{\downarrow}\right]\eq

 We note that this straightforward derivation uses singular value decomposition and does not require any upper bound for $f(A)$. This is an important advantage because it directly shows what the minimum of $g(B)$ should be (which would corresponds to $\min f(A)$ and will be needed for deriving MIN).  A lower bound of GD has been derived in \cite{luofupra10} using only the isometry condition (\ref{conda}). Since we have used more constraints, undoubtedly our bound is sharper .

Before applying this lower bound to solve some interesting related problems, let us show that this bound could be achieved by an infinite number of (collection of measurement-like) operators $\Pi^a=\{|k\ran\lan k|\}$, where each $|k\ran\lan k|$ is a hermitian, unit trace, but not necessarily positive operator. If all $|k\ran\lan k|$ satisfy (\ref{coherentv}), it would correspond to the \emph{optimal} von Neumann measurement $\Pi^a$, which would yield the minimum of GD.  We note that $\Pi^a=\{|k\ran\lan k|\}$ where  \bq\label{vonnumenpi}|k\ran\lan k|=\sum_{i=1}^{m^2}a_{ki}X_i=\frac{1}{m}I_m+\frac{1}{\sqrt{2}}\mathbf{a_k\lambda}, k=1,2,\ldots m-1\eq and $|m\ran\lan m|=I_m-\sum_{k=1}^{m-1}|k\ran\lan k|$. Thus we need to determine only the first $(m-1)$ projections $|k\ran\lan k|$ and for this we should consider only the first $m-1$ rows of $B$. So, denoting corresponding restrictions of $B,\mathbf{e},U,\Sigma$ by $B_{m-1},\mathbf{e}_{m-1},U_{m-1},\Lambda$ respectively, (\ref{restB}) reduces to $B_{m-1}B_{m-1}^t=I_{m-1}-(1-1/m)\mathbf{e}_{m-1}\mathbf{e}_{m-1}^t$. This in turn gives \bq\label{restrestB}B_{m-1}=(\mathbf{a_1},\mathbf{a_2},\ldots,\mathbf{a_{m-1}})^t= U_{m-1}\Lambda V^t\eq where $U_{m-1}$ has $\mathbf{e}_{m-1}$ as its last column, $\Lambda=$diag$(1,1,\ldots,1,1/\sqrt{m})$ and columns of $V$ are the  eigenvectors of $G$ corresponding to eigenvalues $\lambda_k^{\downarrow}$. We note that different choice of $U_{m-1}$  corresponds to different $|k\ran\lan k|$ (though the set $\Pi^a$ may remain invariant). For a particular explicit representation, out of many choices for the rest of the columns, a particular one is to choose $U_{m-1}$ as the Helmert matrix \cite{Helmert} which is given by (for clarity column vectors are not normalized)\bq\label{ukhelmert}U_{m-1}=\left(
                                                                                \begin{array}{cccccc}
                                                                                  1 & 1 & 1 & \ldots & 1 & 1 \\
                                                                                  -1 & 1 & 1 & \ldots & 1 & 1 \\
                                                                                  0 & -2 & 1 & \ldots & 1 & 1 \\
                                                                                  0 & 0 & -3 & \ldots & 1 & 1 \\
                                                                                  0 & 0 & 0 & \ldots & 1 & 1 \\
                                                                                  0 & 0 & 0 & \ldots & -m+2 & 1 \\
                                                                                \end{array}
                                                                              \right)\eq
Denoting the row vectors of $U_{m-1}$ (with normalized columns) as $\mathbf{r'_k}$, we have from (\ref{restrestB}), \bq\label{akfinal}\mathbf{a_k}=\mathbf{r_k}\widetilde{V},\quad k=1,2,\ldots m-1\eq where $\mathbf{r_k}=\mathbf{r'_k}\circ(1,1,\ldots,1,1/\sqrt{m})$ (`$\circ$' is \emph{entrywise} multiplication) and $\widetilde{V}$ is the $m-1\times m^2-1$ left-upper block of $V^t$. We emphasize that for $m\ge4$, the choice $U_{m-1}$ is not unique e.g., for $m=5$, $U_{m-1}$ can be taken as the standard $4\times4$ Hadamard matrix.

\emph{Upper bound for MIN and its saturation by Werner and Isotropic states:}
To calculate MIN for a state $\rho$, we have to find minimum of Tr($ATT^tA^t$) where $A$ has to satisfy an additional constraint (\ref{dmincond}). As in the case of GD, ignoring (\ref{coherentv}) and (\ref{dmincond}) we would get an upper bound of MIN. Setting $G=TT^t$, we see that the required minimum is exactly the minimum of $g(A)$ in (\ref{gB}). Hence just like (\ref{finalg}), we have  \bq\label{finalgmin}\min g(A)=\sum_{k=1}^{m-1}\lambda_{k}^{\uparrow}\eq Thus we have the following upper bound on MIN \bq\label{finalmin}N(\rho)\le\frac{1}{mn}\left[\frac{4}{mn}\|T\|^2-\sum_{k=1}^{m-1}\lambda_k^{\uparrow}\right]=\frac{4}{m^2n^2}\sum_{k=1}^{m^2-m}
\lambda_k^{\downarrow}\eq where $\lambda_k^{\uparrow}$ ($\lambda_k^{\downarrow}$) are the eigenvalues of $TT^t$ sorted in non-decreasing (non-increasing) order. We note that this upper bound is exactly the same as derived in \cite{luofuprl11}. If we set $\mathbf{x}=\mathbf{0}$, the extra constraint (\ref{dmincond}) for MIN gets automatically satisfied. In addition, if all the eigenvalues are equal, the lower bound of $D(\rho)$ in (\ref{finaldiscord}) and
   the upper bound of $N(\rho)$ in (\ref{finalmin}) coincide. So, if one of the bounds saturates, necessarily we will have $D=N$.  As an interesting consequence, we give the following two examples. The $m\times m$ dimensional Werner states $$\rho=\frac{m-z}{m^3-m}\mathbf{1}+\frac{mz-1}{m^3-m}F,\quad z\in[-1,1]$$ with $F:=\sum_{kl}|k\ran\lan l|\otimes |l\ran\lan k|$ has $$D=N=\frac{(mz-1)^2}{m(m-1)(m+1)^2}$$ For the $m\times m$ dimensional isotropic states $$\rho=\frac{1-z}{m^2-1}\mathbf{1}+\frac{m^2z-1}{m^2-1}|\Psi\ran\lan\Psi|,\quad z\in[0,1]$$ with $|\Psi\ran:=1/\sqrt{m}\sum_{k=1}^m|k\ran\otimes|k\ran$ we have $$D=N=\frac{(m^2z-1)^2}{m(m-1)(m+1)^2}$$

\emph{All $2\otimes n$ states saturate our lower bound:} Setting $m=2$, we see from (\ref{ukhelmert}) the unique $U_1$ is just $1$ (seen as $1\times1$ matrix), and hence from (\ref{akfinal}), $\mathbf{a_1}=1/\sqrt{2}\mathbf{v_1}$. Then from (\ref{vonnumenpi}), the unique measurement operators are given by \bqa\label{optimal22pro}|1\ran\lan1|&=&\frac{1}{2}\left(I_2+\mathbf{v_1}\mathbf{\lambda}\right)\nonumber\\|2\ran\lan2|&=&\frac{1}{2}
\left(I_2-\mathbf{v_1}\mathbf{\lambda}\right)\eqa Since $\mathbf{v_1}$ (which is the eigenvector corresponding to the largest eigenvalue of $G$) has norm $1$, both the operators in (\ref{optimal22pro}) are projectors and hence satisfy (\ref{coherentv}). Thus all $2\otimes n$ states saturate our lower bound showing its tightness. We wish to mention that GD for these states have also been derived in \cite{Vinjanampathy}, following the approach of \cite{dakicprl10}.

One immediate consequence of the saturation of lower bound is that it readily gives GD for any $N$ qubit state. This in turn enables us to check monogamy relations etc. for qubit states. We will consider this case in the following paragraph.

\emph{Geometric discord is monogamous for both generalized $GHZ$ and $W$ states of $N$ qubits:--}
Recently many authors have studied monogamy property of different versions of quantum discord \cite{giorgipra11,prabhuar11,disothers}. A correlation measure $\mathcal{Q}$ is said to be monogamous iff for any tripartite state $\rho_{123}$ (generalization to arbitrary state is straightforward) the following inequality  holds \bq\label{dmonogamy} \mathcal{Q}(\rho_{12})+\mathcal{Q}(\rho_{13})\le \mathcal{Q}(\rho_{1|23})\eq The authors of \cite{giorgipra11,prabhuar11} have shown that for (a specific measure of) quantum discord, all 3-qubit pure $W$-type states violate monogamy relation, while the $GHZ$-type states may or may not violate monogamy.  Here we will show that the $N$-qubit generalized $GHZ$ state $|GGHZ\ran=a|00\cdots0\ran+b|11\cdots1\ran$ and the generalized $W$ states $|GW\ran=\sum_{k=1}^N c_k|001_k0\cdots0\ran$ both satisfy monogamy for GD \footnote{Just two days prior to this submission, in an interesting work Streltsov \emph{et.al.} \cite{Streltsovar11} have proven that all pure three qubit states satisfy monogamy of GD.}.

Since GD is non-negative and any bipartite reduced density matrix (RDM) $\rho_{1K}$ of $|GGHZ\ran$ is classical, Eq. (\ref{dmonogamy}) is automatically satisfied for $|GGHZ\ran$. Indeed, the relation holds for any arbitrary Schmidt-decomposable state $\sum \sqrt{\lambda_i}|ii\cdots i\ran$. Thus, GD is monogamous for $|GGHZ\ran$.

In case of $|GW\ran$, being pure, it should have a Schmidt decomposition over the cut $1|23\ldots N$ and the Schmidt coefficients (square-root of eigenvalues of $\rho_1$) are given by $c_1$ and $\sqrt{1-c_1^2}$. Hence by the result of \cite{luofuprl11}, the right hand side of  Eq. (\ref{dmonogamy}) becomes $2$det$(\rho_1)=2c_1^2(1-c_1^2)=2c_1^2(c_2^2+c_3^2+\cdots+c_N^2)$.  To evaluate the left hand side we note that the required bipartite RDMs are given by \bq\label{dgw} \rho_{1k}=\left[
                           \begin{array}{cccc}
                             1-c_1^2-c_k^2 & 0 & 0 & 0 \\
                              0& c_k^2 & c_1 c_k & 0 \\
                              0& c_1 c_k & c_1^2& 0 \\
                              0& 0 & 0 & 0 \\
                           \end{array}
                         \right] \eq

Expressing in Bloch form, we have $\mathbf{x}=(0,0,1-2c_1^2)$ and $T=$diag$(2c_1c_k, 2c_1c_k, 1-2c_1^2-2c_k^2)$. Hence we have by our formula \bq D(\rho_{1k})
=c_1^2c_k^2+\frac{1}{4}\min\{4c_1^2c_k^2,(1-2c_1^2)^2+(1-2c_1^2-2c_k^2)^2\} \eq Using $\min\{a,b\}\le a$, this gives $D(\rho_{1k})\le2c_1^2c_k^2$. Thus, summing over $k$'s our claim follows.

One notable observation is that if we set all $c_k$,s equal ($1/\sqrt{N}$), then Eq. (\ref{dmonogamy}) becomes an equality. This is quite remarkable, because it is known that the same relation holds for the entanglement measure \emph{tangle} $\tau$ \cite{tanglemonogamy}, where the concept of monogamy appeared for the first time.

We will now show that the result remains unchanged even if we add a term $c_0|00\cdots0\ran$ to $|GW\ran$, i.e., if we consider class of states including all $N$-qubit pure states which are equivalent to $W$ states under stochastic local operations and classical communication (SLOCC) \cite{sloccequivalentw}. In this case the right hand side of  Eq. (\ref{dmonogamy}) becomes $2$det$(\rho_1)=2c_1^2(c_2^2+c_3^2+\cdots+c_N^2)$. To evaluate the left hand side, we note that each RDM $\rho_{1k}$ has $\mathbf{x}=(2 c_0 c_1,0,1-2c_1^2)$ and $$T=\left(
\begin{array}{ccc}
 2 c_1 c_k & 0 & 2 c_0 c_1 \\
 0 & 2 c_1 c_k & 0 \\
 2 c_0 c_k & 0 & 1-2 c_1^2-2 c_k^2
\end{array}
\right)$$
Therefore eigenvalues of $\mathbf{xx}^t+TT^t$ are given by $\lambda_1=4 c_1^2 c_k^2$, $\lambda_{2,3}=a\pm \sqrt{b}$ where $a=(1-2c_1^2)^2-2c_k^2(1-c_0^2-c_k^2-c_1^2)+4c_1^2(c_0^2+c_k^2)$ and $b=8 c_1^2 c_k^2 [-(-1+2 c_0^2+2 c_1^2){}^2-2 (-1+3 c_0^2+2 c_1^2) c_k^2-2 c_k^4]+a^2$. Noting that $\|\mathbf{x}\|^2+\|T\|^2=8 c_1^2 c_k^2+8 c_0^2 c_1^2+(1-2 c_1^2){}^2+4 c_0^2 c_k^2+(1-2 c_1^2-2 c_k^2){}^2:=8 c_1^2 c_k^2+c$, we have \bqa\label{monsloccw}\|\mathbf{x}\|^2+\|T\|^2-\max\{\lambda_1,\lambda_2,\lambda_3\}&\le&\|\mathbf{x}\|^2+\|T\|^2-\lambda_2\nonumber\\
&=& 8 c_1^2 c_k^2+c-(a+\sqrt{b})\nonumber\\ &\le& 8 c_1^2 c_k^2+c-a-|c-a|\nonumber\\&\le& 8 c_1^2 c_k^2\eqa where we have used $b=(c-a)^2+32c_0^2c_1^2c_k^2(1-c_0^2-c_1^2-c_k^2)\ge(c-a)^2$. Hence $D(\rho_{1k})\le2c_1^2c_k^2$ and summing over $k$'s the desired result follows.

Due to this similarity with tangle it may be tempting to think that GD is also monogamous (at least) for all $N$-qubit pure states. But GD, in contrast to tangle, is not monogamous for mixed states \cite{Streltsovar11}. This indicates that may be GD is not monogamous for all pure states. To show this, let us consider the following $N$-qubit pure state:\bq\label{4nonmonogd}|\psi\ran=\sqrt{p}|00\cdots0\ran+\sqrt{1-p}|+1\cdots1\ran\eq where $|+\ran=1/\sqrt{2}(|0\ran+|1\ran)$. For this state, we have $D(\rho_{1|23\ldots N})=2$det$(\rho_1)=p(1-p)$, whereas $D(\rho_{1k})=1/2\min\{p^2,(1-p)^2\}$. The state being symmetric in parties $2,3,\ldots,N$, monogamy relation (\ref{dmonogamy}) is satisfied iff \bq\label{finalnonmonogd}\frac{N-1}{2}\min\{p^2,(1-p)^2\}\le p(1-p)\eq
Clearly all $$p\in\left(\frac{2}{N+1}, \frac{N-1}{N+1}\right)$$ violate this relation. Thus not all pure states, beyond $3$-qubits, satisfy monogamy of GD.

To conclude, we have derived in a very simple way, a tight lower bound for geometric discord of arbitrary bipartite states which is saturated by all $2\otimes n$ states. We have also shown that Werner and isotropic states have same amount of geometric discord and measurement induced non locality. All pure $N$-qubit generalized $GHZ$ and $W$ states are shown to satisfy monogamy of geometric discord. Giving an example we have shown that not all pure states of four or higher qubits satisfy monogamy of geometric discord.

We would like to thank P. S. Joag for helpful discussions.

\end{document}